\documentclass[twocolumn, secnumarabic, amssymb, nobibnotes, aps, pra]{revtex4}
\usepackage{graphicx}

\usepackage{epsfig,psfrag}

\begin{document}

\title{Bragg spectroscopy of an accelerating condensate with solitary-wave behaviour}

\author{K J Challis}
\email{kchallis@physics.otago.ac.nz}
\affiliation{Department of Physics, University of Otago, PO Box 56, Dunedin, New Zealand}
\author{R J Ballagh}
\affiliation{Department of Physics, University of Otago, PO Box 56, Dunedin, New Zealand}

\date{\today}

\begin{abstract}
We present a theoretical treatment of Bragg spectroscopy of an accelerating condensate in a solitary-wave state.  Our treatment is based on the Gross-Pitaevskii equation with an optical potential representing the Bragg pulse and an additional external time-dependent potential generating the solitary-wave behaviour.  By transforming to a frame translating with the condensate, we derive an approximate set of equations that can be readily solved to generate approximate Bragg spectra.  Our analytic method is accurate within a well defined parameter regime and provides physical insight into the structure of the spectra.  We illustrate our formalism using the example of Bragg spectroscopy of a condensate in a time-averaged orbiting potential trap.
\end{abstract}

\maketitle

\section{Introduction}

Bragg spectroscopy has proved to be a precise and versatile technique for
measuring a variety of properties of a Bose-Einstein condensate, including the 
momentum distribution \cite{Ketterle99} and coherence functions \cite{Aspect03}.  A theoretical proposal has been made to use Bragg spectroscopy
to provide a signature of a vortex state in a condensate \cite{Blakie01},
and recently the technique has been suggested as a means for characterizing
superfluid pairing in fermionic systems \cite{Zwerger04}.

The theoretical description of Bragg spectroscopy of a condensate is well established and Blakie et al \cite{Blakie02}, for example, have shown that a mean-field treatment using the Gross-Pitaevskii equation is an appropriate formalism.  Unfortunately numerically calculating a Bragg spectrum using the Gross-Pitaevskii equation is highly computationally intensive and may not readily yield a physical interpretation of the system.  

In this paper we present a treatment of Bragg spectroscopy of a condensate which is accelerating in the form of a solitary-wave due to the influence of an external time-dependent trapping potential.  Solitary-wave behaviour occurs only for a certain class of potentials, but is not uncommon.  A simple example is the dipole oscillation of a condensate in a harmonic trap.  Our method is based on transforming to the frame of reference where the condensate is stationary, allowing us to establish a relatively simple set of equations that can be readily solved to give approximate Bragg spectra.  We also obtain an analytic approximation to the spectrum, which provides physical insight into its structure.

The paper is organised as follows.  In Sec.~\ref{sec:spec} we describe our
theoretical formalism, introducing solitary-wave solutions of the Gross-Pitaevskii equation and transforming the equation to the translating
frame.  We outline a simple two-state momentum-space formalism which can be
used to generate approximate Bragg spectra for a general solitary-wave, and we determine the validity regime in which such spectra are accurate.  In Sec.~\ref{sec:TOP} we use the methods of Sec.~\ref{sec:spec} to treat Bragg spectroscopy of a condensate undergoing micromotion in a time-averaged orbiting potential trap. We demonstrate the utility of  our approximate solutions by comparing them with Bragg spectra calculated by simulating the Gross-Pitaevskii equation. Bragg spectroscopy in a time-averaged orbiting potential trap has been investigated
experimentally \cite{Wilson03} and our treatment provides quantitative
agreement with full theoretical calculations and qualitative agreement with
the experimental results \cite{Challis03}.

\section{Bragg spectroscopy of a solitary-wave \label{sec:spec}}

Our theoretical treatment is based on solving the Gross-Pitaevskii equation for a condensate with wave function $\psi(\mathbf{r},t),$ in the presence of a time-dependent external trapping potential and a Bragg pulse, i.e., 
\begin{eqnarray}
\label{GPE}
i \frac{\partial}{\partial t} \psi (\mathbf{r},t) & = & \left(-\nabla ^{2}+V(\mathbf{r},t)+V_{\rm{opt}}(\mathbf{r},t)+ \right. \nonumber \\ & &  \left. C|\psi(\mathbf{r},t)|^2 \right)  \psi (\mathbf{r},t).
\end{eqnarray}
We have chosen to present our discussion in terms of dimensionless units where the time scale is given by the inverse of a characteristic trap frequency $\omega_x$ (e.g., as defined in Sec.~\ref{sec:harm}) and the position scale is $x_0=\sqrt{\hbar/2 m \omega_x}$ (where $m$ is the mass of an atom).  The external trapping potential $V(\mathbf{r},t)$ can be decomposed without loss of generality as
\begin{equation}
V(\mathbf{r},t)=V(\mathbf{r})+G(\mathbf{r},t),
\label{pot}
\end{equation}
and the optical potential generated by the Bragg pulse is 
\begin{equation}
\label{opt_pot-3d}
V_{\rm opt}(\mathbf{r},t)=\frac{1}{2} U_0 \cos (\mathbf{q}\cdot\mathbf{r}-\omega t),
\end{equation}
where $\mathbf{q}$ and $\omega$ are, respectively, the wave-vector difference and the frequency difference between the two laser beams forming the Bragg pulse \cite{Blakie00}.  The dimensionless nonlinearity coefficient $C$ is defined in terms of the number of condensed atoms $N$, and the $s$-wave scattering length $a,$ i.e.,
\begin{equation}
C=\frac{4\pi\hbar a N}{m\omega_x x_0^3}.
\end{equation}

\subsection{Solitary-wave solutions \label{sec:solitary}}

Solitary-wave solutions to the Gross-Pitaevskii equation in the absence of the optical potential [Eq.~(\ref{GPE}) with $U_0=0$] exist for particular forms of the external potential $V(\mathbf{r},t)$ of Eq.~(\ref{pot}).  Such solutions, which evolve in three dimensions without changing shape, have been discussed previously \cite{Morgan97,Margetis99,Challis04} and have the form
\begin{equation}
\label{psi-postulate}
\psi_{\rm{SW}}(\mathbf{r},t)=\xi(\mathbf{r}-\bar{\mathbf{r}}(t))e ^{-i \mu t+i S(\mathbf{r},t)},
\end{equation}
where the envelope wave function $\xi(\mathbf{r})$ is an eigenstate of the time-independent Gross-Pitaevskii equation with the time-independent trapping potential $V(\mathbf{r})$ and chemical potential $\mu,$ i.e.,
\begin{equation}
\label{GPE-phi}
\mu \xi (\mathbf{r})=\left( -\nabla ^{2}+V(\mathbf{r})+C|\xi(\mathbf{r})|^2 \right) \xi (\mathbf{r}).
\end{equation}
The position offset in the envelope wave function is
\begin{equation}
\bar{\mathbf{r}}(t) = \int \psi^*_{\rm{SW}}(\mathbf{r},t) \mathbf{r} \psi_{\rm{SW}}(\mathbf{r},t) d \mathbf{r} -\int \xi^*(\mathbf{r}) \mathbf{r} \xi(\mathbf{r}) d \mathbf{r}.
\end{equation}
The phase $S(\mathbf{r},t)$ is determined by substituting the solitary-wave solution (\ref{psi-postulate}) into the time-dependent Gross-Pitaevskii equation \cite{Challis04}.  We restrict our discussion to solutions where the velocity of the envelope $\xi(\mathbf{r}-\bar{\mathbf{r}}(t))$ is proportional to the gradient of $S(\mathbf{r},t)$ [in our units $2 \nabla S(\mathbf{r},t)$], and equating this to the centre of mass velocity $d \bar{\mathbf{r}}(t)/dt$ yields
\begin{equation}
S(\mathbf{r},t)=\frac{1}{2} \mathbf{r} \cdot \frac{d \bar{\mathbf{r}}(t)}{dt}+\frac{1}{4}\int  \left[ \bar{\mathbf{r}}^2(t)-\left( \frac{d \bar{\mathbf{r}}(t)}{dt} \right)^2 \right] dt.
\label{Sfunc}
\end{equation}
The centre of mass motion of the solitary-wave solutions can be shown to obey a form of Ehrenfest's theorem,
\begin{equation}
\frac{1}{2}\frac{\partial ^2 \bar{\mathbf{r}}(t)}{\partial t^2}=-\nabla F (\mathbf{r},t),
\label{Ehrenfest}
\end{equation}
where 
\begin{equation}
F(\mathbf{r},t)=V(\mathbf{r},t)-V(\mathbf{r}-\bar{\mathbf{r}}(t)).
\label{rest}
\end{equation}
For solitary-wave solutions to exist the function $F(\mathbf{r},t)$ must yield a solution to Eq.~(\ref{Ehrenfest}) and, therefore, can be at most linear in $\mathbf{r}.$

\subsubsection{Harmonic trap \label{sec:harm}}

In Bose condensed systems it is most common that the time-independent trapping potential $V(\mathbf{r})$ is harmonic, i.e.,
\begin{equation}
V_{\rm H}(\mathbf{r})=\frac{1}{4}(x^2+y^2+\lambda^2z^2),
\end{equation}
with frequency $\omega_x$ in the $x$-$y$ plane (in SI units), aspect ratio $\lambda,$ and where $\mathbf{r}=(x,y,z)$.  In the absence of the Bragg pulse ($U_0=0$), the Gross-Pitaevskii equation (\ref{GPE}) has solitary-wave solutions  provided that $G(\mathbf{r},t)$ has the particular form  such that Eq.~(\ref{Ehrenfest}) is satisfied, i.e.,
\begin{equation} 
G(\mathbf{r},t)=\mathbf{g}_1(t) \cdot \mathbf{r}+g_2 (t).
\label{P_harm}
\end{equation}

\subsection{Translating frame \label{sec:frame}}

We begin our treatment of Bragg spectroscopy in a time-dependent external trapping potential by transforming to the frame translating such that the condensate centre of mass is stationary in that frame.  While numerical solutions of the Gross-Pitaevskii equation can be calculated in the lab frame, the description of the system in the translating frame, where the condensate is at rest, allows simple approximate methods and a physical interpretation to be developed since the only time-dependence for the system in the translating frame appears via an effective optical potential.

A general quantum mechanical transformation to an accelerating translating frame has been given earlier \cite{Challis04}.  For the particular case we are interested in, of solitary-wave motion, the translation in co-ordinate space is defined by setting the coordinates of the new frame to be
\begin{equation}
\mathbf{R} = \mathbf{r}-\bar{\mathbf{r}}(t),
\label{trans}
\end{equation}
where we denote the components of $\mathbf{R}$ to be $(X,Y,Z).$  The momentum in the translating frame is derived by differentiating Eq.~(\ref{trans}) yielding
\begin{equation}
\mathbf{P}  = \mathbf{p} -\frac{1}{2} \frac{d\bar{\mathbf{r}}(t)}{dt},
\label{trans_p}
\end{equation}
where we have used the fact that in our dimensionless units $\mathbf{p}=\mathbf{v}/2$. The wave function in the translating frame is
\begin{equation}
\psi ^{\rm{t}} (\mathbf{R},t)=e ^{-\frac{1}{2} i \mathbf{R} \cdot \frac{d \bar{\mathbf{r}}(t)}{dt}}\psi (\mathbf{r},t),
\label{int_picture}
\end{equation}
facilitating the momentum translation of Eq.~(\ref{trans_p}) \footnote{If $\bar{\mathbf{r}}(t) \cdot d\bar{\mathbf{r}}(t)/dt \neq 0$ the unitary operators that generate the position and momentum translations of Eqs. (\ref{trans}) and (\ref{trans_p}) may not commute but the additional factor in the unitary transformation to an accelerating translating frame is a time-dependent phase which can be removed in an interaction picture.}.  The Gross-Pitaevskii equation in the translating frame can now be derived by transforming Eq.~(\ref{GPE}) using Eqs.~(\ref{trans}) and (\ref{int_picture}) to yield
\begin{eqnarray}
\label{GPE-trans}
i \frac{\partial}{\partial t} \psi^{\rm t} (\mathbf{R},t) & = & \left(-\nabla_{\mathbf{R}} ^{2}+W^{\rm t} (\mathbf{R},t)+V^{\rm t}_{\rm{opt}}(\mathbf{R},t)+ \right. \nonumber \\ & &  \left. C|\psi^{\rm t}(\mathbf{R},t)|^2 \right)  \psi ^{\rm t} (\mathbf{R},t).
\end{eqnarray}
In Eq.~(\ref{GPE-trans}) we have denoted 
\begin{equation}
\nabla _{\mathbf{R}}=\left( \frac{\partial}{\partial X},\frac{\partial}{\partial Y},\frac{\partial }{\partial Z} \right),
\end{equation}
and the optical potential in the translating frame is 
\begin{equation}
\label{opt_pot-3d-trans}
V^{\rm t}_{\rm opt}(\mathbf{R},t)=\frac{1}{2} U_0 \cos (\mathbf{q}\cdot\mathbf{R}+\mathbf{q}\cdot \bar{\mathbf{r}}(t)-\omega t).
\end{equation}
The potential denoted $W^{\rm t} (\mathbf{R},t)$ is the sum of the external trapping potential in the translating frame and additional terms due to the momentum translation, i.e.,
\begin{equation}
W^{\rm t} (\mathbf{R},t)=V(\mathbf{R}+\bar{\mathbf{r}}(t),t)+\frac{1}{2}\mathbf{R}\cdot \frac{d^2 \bar{\mathbf{r}}(t)}{dt^2}-\frac{1}{4}\left( \frac{d\bar{\mathbf{r}}(t)}{dt} \right) ^2.
\label{pot-trans}
\end{equation}
By taking the gradient of Eq.~(\ref{pot-trans}) and making a comparison with Eq.~(\ref{Ehrenfest}) we find that 
\begin{equation}
\nabla _{\mathbf{R}} W^{\rm t} (\mathbf{R},t) = \nabla _{\mathbf{R}} V(\mathbf{R})
\end{equation}
and, therefore,
\begin{equation}
W^{\rm t} (\mathbf{R},t)=V(\mathbf{R})+A(t),
\end{equation}
where $A(t)$ is a function only of time.  Defining an interaction picture by 
\begin{equation}
\psi ^{\rm T} (\mathbf{R},t)=\psi ^{\rm t} (\mathbf{R},t) e^{i \int A(t)dt},
\end{equation}
the Gross-Pitaevskii equation becomes 
\begin{eqnarray}
\label{GPE-trans-T}
i \frac{\partial}{\partial t} \psi^{\rm T} (\mathbf{R},t) & = & \left(-\nabla_{\mathbf{R}} ^{2}+V(\mathbf{R})+V^{\rm t}_{\rm{opt}}(\mathbf{R},t)+ \right. \nonumber \\ & &  \left. C|\psi^{\rm T}(\mathbf{R},t)|^2 \right)  \psi ^{\rm T} (\mathbf{R},t),
\end{eqnarray}
which shows that in the translating frame all the explicit time-dependence of the system is contained within the optical potential term [see Eq.~(\ref{opt_pot-3d-trans})].  When the Bragg pulse is turned off, stationary solutions of Eq.~(\ref{GPE-trans-T}) exist of the form
\begin{equation}
\psi _{\rm SW}^{\rm T} (\mathbf{R},t)=\xi (\mathbf{R})e^{-i \mu t},
\end{equation}
which are solitary-wave solutions in the lab frame with centre of mass motion governed by Eq.~(\ref{Ehrenfest}).

\subsection{Simple momentum formalism \label{sec:mom}}

A theoretical formalism for describing Bragg spectroscopy of a stationary condensate has been developed by Blakie et al.~\cite{Blakie00}.  Those authors derived an analytic treatment of Bragg scattering, for the case where $C=0$, using an approximate two-state momentum space model to consider condensates released from a harmonic trap at the instant that the Bragg pulse is applied.  Their treatment provides a simple approximate method for generating Bragg spectra which, within a given parameter regime, are in quantitative agreement with full calculations of the Gross-Pitaevskii equation.  Here we extend that model to describe Bragg spectroscopy of condensates moving with solitary-wave behaviour.

Following broadly the approach of Blakie et al.~\cite{Blakie00} we place three restrictions on the system.  First, mean-field effects are neglected by taking $C=0.$  Our treatment of the nonlinear term is detailed in Sec.~\ref{sec:dim}.  Secondly, the time-independent trapping potential $V(\mathbf{r})$ is neglected while retaining the time-dependent $G(\mathbf{r},t)$.  This means that the condensate centre of mass motion is still governed by Eq.~(\ref{Ehrenfest}) but the condensate is no longer confined about its centre of mass.  When neglecting atomic interactions, removing the trap $V(\mathbf{r})$ does not alter the momentum distribution of the unscattered condensate, whereas the condensate part scattered by the Bragg pulse is decelerated.  Provided that the Bragg pulse length is short compared to the oscillation period of atoms confined by $V(\mathbf{r})$,  the effect of the confining potential on the Bragg spectrum is negligible.  In some experimental systems it may be possible to apply $V(\mathbf{r})$ and $G(\mathbf{r},t)$ independently so that $V(\mathbf{r})$ can be turned off when the Bragg pulse is applied.  Thirdly, for simplicity, we choose the wave vector $\mathbf{q}$ of the optical potential to lie along the $x$ axis such that $\mathbf{q} \cdot \mathbf{R}=qX$ and the optical potential of Eq.~(\ref{opt_pot-3d-trans}) becomes
\begin{equation}
V_{\rm opt}^{\rm t} (X,t) = \frac{1}{2} U_0 \cos (q X+q \bar{x} (t)-\omega t).
\label{opt-pot-trans-1d}
\end{equation}
This can be done without loss of generality for the case where $V(\mathbf{r})$ is turned off.  Finally then, within these restrictions, the wave function $\psi^{\rm T}(\mathbf{R},t)$ is seperable and denoting the product wave function in $X$ by $\psi^{\rm T}(X,t)$ it is possible to reduce the analysis of the system to one dimension without loss of generality \cite{Blakie00}.  

We begin by transforming the one-dimensional equivalent of Eq.~(\ref{GPE-trans-T}), with $C=0$ and $V(\mathbf{R})=0,$ into momentum space using the Fourier integral 
\begin{equation}
\label{Fint}
\phi^{\rm T} (K,t)=\frac{1}{\sqrt{2\pi}}\int \psi^{\rm T} (X,t) e^{-iK X}dX,
\end{equation}
to give
\begin{eqnarray}
i\frac{\partial }{\partial t} \phi ^{\rm T} (K,t)& = & K^{2} \phi ^{\rm T}(K,t) \nonumber \\ & &
 +\frac{1}{4} U_0 \left[ e^{-i(\omega t-q \bar{x}(t))}\phi ^{\rm T}(K-q,t) \right. \nonumber \\
& & \left. +e^{i(\omega t-q\bar{x}(t))}\phi^{\rm T} (K+q,t)\right].
\label{momentum-space GPE} 
\end{eqnarray}
Next, we partition momentum space into bins of width $q$, centred around the Bragg orders $K=nq$, where $n$ is integer.  That is achieved by expressing the wave function $\phi^{\rm T} (K,t)$ as a set of wave functions $\tilde{\phi}^{\rm T}_n(\tilde{K},t)$, each defined only within bin $n$ of the partitioned momentum space, i.e.,
\begin{equation}
\tilde{\phi}^{\rm T}_n(\tilde{K},t)=\phi^{\rm T}(K,t),
\end{equation}
where
\begin{equation}
\tilde{K} : \left( n-\frac{1}{2} \right) q<K< \left( n+\frac{1}{2} \right) q.
\end{equation}
The description of the system is simplified by defining $\kappa=\tilde{K} -nq$ and writing 
\begin{equation}
\phi^{\rm T}_n(\kappa,t)=\tilde{\phi}^{\rm T}_n(\tilde{K},t) e^{in(\omega t-q\bar{x}(t))},
\end{equation}
where $\kappa$ varies from $-q/2$ to $q/2$, for every bin.  The momentum space Schr$\rm{\ddot{o}}$dinger equation can then be written as an infinite set of equations coupling wave functions of consecutive bins, 
\begin{eqnarray}
i\frac{\partial }{\partial t} \phi^{\rm T} _{n}(\kappa ,t)& = & \omega _n (\kappa,t) \phi ^{\rm T}_n (\kappa ,t) \nonumber \\
& & +\frac{1}{4} U_{0} \left[ \phi ^{\rm T} _{n-1}(\kappa ,t)+\phi ^{\rm T}_{n+1}(\kappa ,t)\right],
\label{mom_coupled}
\end{eqnarray}
where
\begin{equation}
\label{freq_defn}
\omega _n (\kappa,t)=(\kappa+nq)^2-n\omega+n q \frac{d\bar{x}(t)}{dt}.
\end{equation}
Equation (\ref{mom_coupled}) is identical to that derived by Blakie et al.~\cite{Blakie00} for a stationary untrapped condensate, with the exception of an additional time-dependence appearing in $\omega _n (\kappa,t).$  The additional term $nqd\bar{x}(t)/dt$ is due to the momentum of the condensate in the lab frame [see Eq.~(\ref{trans_p})] and has previously been referred to as the Doppler term \cite{Wilson03}.

To determine the Bragg spectrum of an accelerating condensate we calculate the population in the first Bragg order ($n=+1$ bin) as a function of the frequency difference $\omega$ in the optical potential.  In the translating frame this is given by
\begin{equation}
P_{+1}(\omega,t) = \int _{-\frac{1}{2}q} ^{\frac{1}{2}q} |\phi_{+1} ^{\rm T}(\kappa,t)|^2   d\kappa,
\end{equation}
corresponding in the lab frame to 
\begin{equation}
P_{+1}(\omega,t)  =  \int _{\frac{1}{2} \left( q+\frac{d\bar{x}(t)}{dt} \right) } ^{\frac{1}{2} \left( 3q+\frac{d\bar{x}(t)}{dt} \right) } |\phi (k,t)|^2   dk,
\label{popn_lab}
\end{equation}
where $\phi (k,t)$ is defined by
\begin{equation}
\label{Fint_lab}
\phi (k,t)=\frac{1}{\sqrt{2\pi}}\int \psi (x,t) e^{-ikx}dx
\end{equation}
in terms of the wave function in the lab frame $\psi (x,t)$, and the limits of integration in Eq.~(\ref{popn_lab}) account for the solitary-wave centre of mass motion of the condensate.  

Solving Eq.~(\ref{mom_coupled}), in order to calculate $\phi_{+1} ^{\rm T}(\kappa,t),$ is no more straight-forward than solving the full Schr$\rm{\ddot{o}}$dinger equation.  Again following the treatment of Blakie et al.~\cite{Blakie00} the problem becomes readily tractable if we assume that in the region of interest, where first order Bragg scattering is strong, only the $n=0$ and $n=+1$ momentum wave functions are substantially populated.  In this `two-state model' Eq.~(\ref{mom_coupled}) reduces to 
\begin{eqnarray}
i\frac{\partial }{\partial t}\left[ \begin{array}{c}
\phi ^{\rm T}_{0}(\kappa ,t)\\
\phi ^{\rm T}_{+1}(\kappa ,t)
\end{array}\right] \hspace{3cm} & &  \nonumber \\
= \left[ \begin{array}{cc}
\omega _{0}(\kappa) & \frac{1}{4} U_0 \\ 
\frac{1}{4} U_0 & \omega _{0}(\kappa)+\Delta_{0\rightarrow +1} (\kappa,t)
\end{array}\right] & \left[ \begin{array}{c}
\phi^{\rm T} _{0}(\kappa ,t)\\
\phi^{\rm T}_{+1}(\kappa ,t)
\end{array}\right],
\label{two_state_eqn}
\end{eqnarray}
where $\omega_0(\kappa)=\kappa^2$ and $\Delta_{0\rightarrow +1} (\kappa,t)=q^2+2\kappa q-\omega+qd\bar{x}(t)/dt.$  Equation (\ref{two_state_eqn}) can be solved easily by numerical means and its solution accurately describes the system when $n\neq0,+1$ populations are negligible.  The validity conditions for Eq.~(\ref{two_state_eqn}) are now derived.

\subsubsection{Validity of the two-state model}

The two-state model is valid when only the $n=0$ and $n=+1$ bins are significantly populated.  Accounting for two-photon transitions only (which couple consecutive bins) the detuning for the transition from bin $n$ to bin $m$ is defined by 
\begin{equation}
\Delta_{n\rightarrow m} (\kappa,t)=\omega_m (\kappa,t)-\omega_n(\kappa,t),
\end{equation}
where $|m-n|=1$.  We define a transition to be on resonance when the magnitude of the detuning is less than $U_0/2$ (determined by the power-broadened width \cite{Blakie00}).  Thus, the first order Bragg transition (between the $n=0$ and $n=+1$ bins) will be resonant for some part of the initial condensate momentum distribution over the frequency range 
\begin{equation}
\omega > q^2+q \left.\frac{d \bar{x}(t)}{dt}\right|_{\rm min}-\sigma q-\frac{1}{2}U_0
\label{freq_a}
\end{equation}
and 
\begin{equation}
\omega < q^2+q \left. \frac{d \bar{x}(t)}{dt}\right|_{\rm max}+\sigma q+\frac{1}{2}U_0,
\label{freq_b}
\end{equation}
where $\sigma$ is the full momentum width of the condensate.  We have chosen to use the $1/e$ condensate momentum width and the extreme values of $d \bar{x}(t)/dt$ in order to yield the largest possible transition width and, therefore, the strictest condition for the two-state validity.  Within the frequency range defined by Eqs.~(\ref{freq_a}) and (\ref{freq_b}) we require that the transitions to neighbouring bins ($n=-1$ and $n=+2$) are not resonant, i.e., $|\Delta_{-1\rightarrow 0} (\kappa,t)| > U_0/2$ and $|\Delta_{+1\rightarrow +2} (\kappa,t)| > U_0/2$.  Those conditions are satisfied when
\begin{equation}
U_0 < 2q (q-\sigma-\delta_p),
\label{2_state_validity}
\end{equation}
where $2 \delta_p=d\bar{x}(t)/dt|_{\rm{max}}-d\bar{x}(t)/dt|_{\rm{min}}.$  Note that Eq.~(\ref{2_state_validity}) requires $\sigma<q,$ which also insures that momentum wave functions of consecutive bins do not overlap.

The discussion leading to Eq.~(\ref{2_state_validity}) does not account for higher order processes.  It is possible for $2m-$photon processes to transfer population from the $n=0$ to the $n=m$ bin, without intermediate bins accumulating significant population.  However, provided the Rabi period for such processes is significantly longer than the Bragg pulse length, the population transfer is negligible and will not effect the two-state dynamics between the $n=0$ and $n=+1$ bins.  From the higher order processes, the four-photon processes are the most likely to effect the two-state validity.  In the case of no centre of mass motion ($\bar{x}(t)=0$ for all $t$), the Rabi frequency at the resonance of the $n=0$ to $n=+2$ transition is \cite{Blakie03}
\begin{equation}
\Omega_{{\rm Rabi} (0\rightarrow +2)}=\frac{U_0^2}{8(\omega_{0}(0)-\omega_{+1}(0))}=\frac{U_0^2}{8q^2}.
\end{equation}
Complete population transfer (for $\kappa=0$) would occur for $\Omega_{{\rm Rabi} (0\rightarrow +2)}t = \pi,$ yielding the additional condition
\begin{equation}
t \ll \frac{8 \pi q^2}{U_0^2}.
\label{2_state_validity_again}
\end{equation}
The condition derived equivalently on the $n=-1$ to $n=+1$ transition is more lenient and therefore, provided that Eqs.~(\ref{2_state_validity}) and (\ref{2_state_validity_again}) are satisfied, all other transitions can be neglected and the two-state approach is valid.

\subsubsection{Simplified treatment of the two-state model:  Main features of the Bragg spectrum \label{sec:features}}

The two-state model provides approximate Bragg spectra for a condensate with solitary-wave behaviour and requires much less computational resources than calculations of the full Gross-Pitaevskii equation.  We will demonstrate the success of the two-state solutions to Eq.~(\ref{two_state_eqn}) in Sec.~\ref{sec:spectra_TOP}.  However, the dominant features of the Bragg spectrum can be readily obtained from a particular simplification of the two-state model, which we outline in this section. 

Blakie et al.~\cite{Blakie00} have shown that if the optical potential~(\ref{opt_pot-3d-trans}) has the simple form $b \cos (\mathbf{q}\cdot \mathbf{R}-\omega t)/2,$ then for the case $V(\mathbf{R})=0$ and $C=0,$ the detuning in the two-state model becomes time-independent, i.e.,
\begin{equation}
\Delta_{0\rightarrow +1} (\kappa)=q^2+2\kappa q-\omega.
\label{delta_tind}
\end{equation}
In that case, the analytic solution to Eq.~(\ref{two_state_eqn}) yields the Bragg spectrum
\begin{equation}
P_{+1} (\omega, t)=\int_{-\frac{1}{2}q} ^{\frac{1}{2} q} \frac{b^2|\phi_0 (\kappa,0)|^2}{8\Omega _{\rm Rabi}^2 (\kappa)}  [1-\cos \Omega _{\rm Rabi} (\kappa) t] d\kappa,
\label{spectrum}
\end{equation}
where the Rabi frequency is 
\begin{equation}
\Omega _{\rm Rabi} (\kappa) =\sqrt{\left(\frac{b}{2}\right)^2+\Delta_{0\rightarrow +1}^2 (\kappa)}.
\label{rabifreq}
\end{equation}
The frequency width of the first order Bragg transition is 
\begin{equation}
\sigma_{\omega} = \sqrt{b^2+\left(\frac{\pi}{t} \right)^2},
\label{lin_width}
\end{equation}
where contributions from power broadening and temporal broadening are included.  In the limit $b t/4 \ll 1$ the total scattered population across the single Bragg resonance is 
\begin{equation}
\int^{\infty} _{-\infty} P_{+1}(\omega,t)d\omega=\frac{\pi b^2  t}{8}.
\label{popn_out_total}
\end{equation}

For the more general case we are considering here, the optical potential in the translating frame is given in one dimension by Eq.~(\ref{opt-pot-trans-1d}), but it can be expanded as a linear sum of simple Bragg potentials with the form considered by Blakie et al.~\cite{Blakie00}, i.e.,
\begin{equation}
V^{\rm t}_{\rm opt} (X,t)=\frac{1}{2}\sum _{l} c_l \cos (qX-(\omega +\omega_l) t+\epsilon _l).
\label{expans_pot}
\end{equation}
The strength, relative phase, and resonance position for each term in the sum can be determined by finding $c_l,$ $\epsilon_l,$ and $\omega_l,$ respectively.  Those parameters can be calculated relatively easily by equating Eqs.~(\ref{opt-pot-trans-1d}) and (\ref{expans_pot}), decomposing the former into trigonometric functions, and comparing terms.  This leads to 
\begin{equation}
U_0 e^{i q \bar{x} (t)}=\sum_l c_l e^{-i\omega_l t+i\epsilon_l},
\label{totrans}
\end{equation}
and taking the fourier transform of both sides of Eq.~(\ref{totrans}) we find that 
\begin{equation}
U_0 \int_{-\infty}^{\infty}e^{i q \bar{x}(t)}e^{-i \varpi t} dt =   \sum_l c_l e^{i \epsilon_l} \delta (\varpi+\omega_l),
\end{equation}
which allows $c_l,$ $\epsilon_l,$ and $\omega_l$ to be readily extracted, for all~$l$.  

The Bragg spectrum of a condensate accelerating as a solitary-wave will have a number of resonance peaks due to the multiple Bragg terms in the linear sum of Eq.~(\ref{expans_pot}).  When the individual Bragg spectra due to consecutive terms in frequency space overlap, the effect due to simultaneous application of the Bragg fields can be complicated.  However, there are particular regimes where the Bragg response at each $\omega_l$ can be considered as if it were acting independently of the others, and then Eqs.~(\ref{delta_tind}) to (\ref{popn_out_total}) can be applied to each resonance, and it is straight-forward to construct a complete approximate spectrum by summing the Bragg spectra due to each individual term.  We refer to this method of calculating approximate spectra as the {\it independent resonance approach}.  The most important case where this approach applies is when the position of consecutive resonances are separated by more than the widths of those resonances, i.e.,
\begin{equation} \omega_{l+1}-\omega_l>\frac{\sigma_{\omega}(c_{l+1})+\sigma_{\omega}(c_l)}{2}.
\label{inter_sep}
\end{equation}
Other particular cases arise, for example in the perturbative limit $c_l t/4 \ll 1,$ the independent resonance approach is accurate when 
\begin{equation}
\cos \left[ \frac{1}{2}(\omega_{l+1}-\omega_l)t-(\epsilon_{l+1}-\epsilon_l) \right]=0.
\label{inter_off}
\end{equation}

\subsection{Nonlinear effects\label{sec:dim}}

Calculating numerical solutions to the three-dimensional time-dependent Gross-Pitaevskii equation is computationally intensive.  Blakie et al.~\cite{Blakie00} demonstrated that within a well defined parameter regime the analytic results derived for the case where $C=0$ provide a good representation of the main behaviour observed in the many-body system.  For the systems we are considering, where the condensate has solitary-wave behaviour in the absence of the Bragg pulse, the centre of mass motion of the condensate is independent of the nonlinear strength of the system and, therefore, in the translating frame the validity regime definitions derived by Blakie et al.~\cite{Blakie00} apply.

There are three dominant many-body effects on Bragg scattering.  The first is that atomic interactions within the condensate cause the momentum distribution to broaden during free expansion.  The time scale for this has been derived to be $t=7\sqrt{3}/4\mu$ \cite{Blakie00} and provided that the Bragg pulse length is shorter than this time scale, the effect of expansion is negligible.  The second many-body effect is a shift in the resonance condition of the first Bragg order due to the mean-field.  For condensates confined in a three-dimensional harmonic trap the resonant frequency is higher by approximately $4\mu/7$ \cite{Blakie00} and this shift can be neglected when $\mu \ll U_0/2.$  In addition to these effects, harmonically trapped condensates with a higher chemical potential have a narrower momentum distribution.  This arises due to the form of $\xi (\mathbf{R}),$ defined by Eq.~(\ref{GPE-phi}):  a broader spatial distribution leads to a narrower momentum distribution.  The Bragg spectra for trapped condensates with a higher chemical potential exhibit narrower spectral features and increased peak scattering.  These effects were also observed by Blakie et al.~\cite{Blakie00}.

In the case where $C=0$ a description of the three-dimensional system can often be simplified to one dimension without loss of generality, as described in Sec.~\ref{sec:mom}.  The Gross-Pitaevskii equation, while it is not separable, can also be reduced to a one-dimensional form (e.g., \cite{Abdullaev03,Salasnich04,Jackson98}).  Bragg spectra calculated using a one-dimensional treatment along the axis of the Bragg pulse also generates reasonably accurate results for a nonlinear condensate, as we will see for a particular example in Sec.~\ref{sec:TOP_nonlinear}.  

\section{Time-averaged orbiting potential trap \label{sec:TOP}}

In this section we calculate Bragg spectra of a condensate accelerating as a solitary-wave.  The spectra are calculated at various levels of approximation: (i) using the approximate methods described in section Sec.~\ref{sec:mom}, (ii) using the full Schr$\rm{\ddot{o}}$dinger equation (Gross-Pitaevskii equation with $C=0$), and (iii) using the Gross-Pitaevskii equation in both one and two dimensions.  

Condensates confined by a harmonic trap will exhibit solitary-wave behaviour for an external potential satisfying Eq.~(\ref{P_harm}).  Some examples include the dipole oscillation of a condensate in a harmonic trap, and the motion resulting from translations of the trap due to the effect of noise.  We have chosen the solitary-wave micromotion of a condensate in a time-averaged orbiting potential (TOP) trap to illustrate our methods.  In this particular system the condensate moves in a circular trajectory following the bias field that removes the discontinuity in the trapping potential (\cite{Challis04} and references therein).  The micromotion velocity of the solitary-wave is so high that the structure of the Bragg spectrum can be measured experimentally with moderate frequency resolution.  The experimental investigation of Bragg spectra in a TOP trap \cite{Wilson03} shows qualitative agreement between the experimental data and calculations made using the one-dimensional Schr$\rm{\ddot{o}}$dinger equation.  The application of our methods of Sec.~\ref{sec:spec} provide the theoretical framework for the concepts introduced previously in relation to Bragg spectroscopy in a TOP trap \cite{Wilson03, Arimondo02}.

\subsection{Theoretical description of the TOP trap}

The TOP trap potential can be well approximated by
\begin{equation}
V_{\rm TOP} (\mathbf{r},t)=V_{\rm H} (\mathbf{r})+r_0 (\cos\Omega t,\sin\Omega t, 0) \cdot \mathbf{r},
\label{TOP_pot}
\end{equation}
where $r_0$ is the so-called `circle of death' and $\Omega$ is the bias field rotation frequency \cite{Challis04}.  Equation (\ref{TOP_pot}) is of the form required for solitary-wave solutions to exist [see Eq.~(\ref{P_harm})] and in fact condensation occurs into the system ground state, which is a solitary-wave solution of the Gross-Pitaevskii equation with circular centre of mass motion given by
\begin{equation}
\bar{\mathbf{r}}(t)=\gamma_{\rm t} (\cos \Omega t, \sin\Omega t,0),
\end{equation}
where $\gamma_{\rm t}=2r_0/(\Omega^2-1)$ \cite{Challis04}.  

\subsection{Effective optical potential}

The solitary-wave description of condensate micromotion allows Bragg spectroscopy in a TOP trap to be described using the theory presented in Sec.~\ref{sec:spec}.  In the translating frame the optical potential of the Bragg pulse becomes, in one dimension,
\begin{equation}
V^{\rm t}_{\rm opt} (X,t) = \frac{1}{2} U_0 \cos (qX+q\gamma_{\rm t}\cos\Omega t-\omega t).
\label{Vopt_trans}
\end{equation}
The optical potential of the Bragg pulse, in the moving frame of the condensate, has been identified previously as being a frequency modulated potential \cite{Wilson03,Arimondo02}.  Our discussion in the translating frame provides a theoretical validation of that identification.

In anticipation of interpreting Bragg spectra of a condensate in a TOP trap we decompose the optical potential in the translating frame in the form of Eq.~(\ref{expans_pot}).  For the frequency modulated potential of Eq.~(\ref{Vopt_trans}) such an expansion is well known, i.e.,
\begin{eqnarray}
V^{\rm t}_{\rm opt}(X,t) & = & \frac{1}{2}U_0 \left\{ J_0 (\gamma q)\cos (qX-\omega t) \right. \nonumber \\
		& &  -\sum^{\infty}_{l=0} (-1)^l J_{2l+1}(\gamma q)
                 \left[ \cos \left( qX-\omega t\right. \right. \nonumber \\
                 & & \left. \left. -(2l +1)\Omega t-\frac{\pi}{2} \right) \right. \nonumber \\
                 & & \left. +\cos \left( qX-\omega t+(2l+1)\Omega t-\frac{\pi}{2} \right) \right] \nonumber \\
                 & & +\sum^{\infty}_{l=1} (-1)^l J_{2l}(\gamma q)
                 \left[ \cos (qX-\omega t-2l\Omega t) \right. \nonumber \\
                 & & \left. \left.  +\cos (qX-\omega t+2l\Omega t) \right] \right\},
\label{Veffect}
\end{eqnarray} 
where $J_{\nu}(z)$ is the Bessel function of the first kind of order $\nu$ and with argument $z$.  In the translating frame, the condensate is subjected to an optical field which is the sum of simple Bragg terms, separated in frequency by the TOP trap bias field rotation frequency.  Therefore, we expect that the Bragg spectra of a condensate with micromotion will have multiple resonances separated by~$\Omega$.

\subsection{Bragg spectra \label{sec:spectra_TOP}}

\begin{figure}
\includegraphics[width=8.5cm]{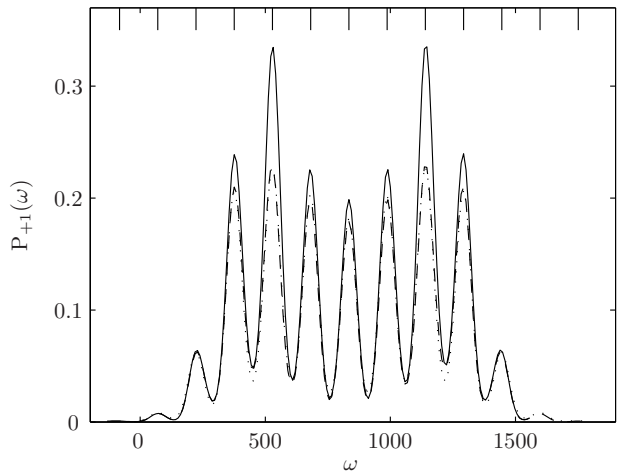}
\caption{\label{fig:TOP_spec} Bragg spectrum of a condensate in a TOP trap calculated using the full Schr$\rm{\ddot{o}}$dinger equation (--), the two-state model (- -), and the independent resonance approach ($\cdot \cdot$).  The vertical lines along the top of the figure indicate the positions of the spectral resonances predicted by Eq.~(\ref{Veffect}).  Parameters are $U_0=45$, $q=29,$ $r_0=1241$, $\Omega =153$, and $t=0.105\pi$ (8~TOP trap cycles).}
\end{figure}

The Bragg spectrum of a condensate in a TOP trap is presented in Fig.~\ref{fig:TOP_spec} for a parameter set which is experimentally accessible \cite{Wilson03}.  The spectrum calculated using the full Schr$\rm{\ddot{o}}$dinger equation (i.e., $C=0$) is shown as a solid line.
The Bragg spectrum is centred at the first order Bragg resonance [see Eqs.~(\ref{freq_a}) and (\ref{freq_b})] and exhibits additional structure compared to the single resonance peak at $\omega=q^2$ which would be expected from a single Bragg term.  Equation~(\ref{Veffect}) correctly predicts the positions of the resonance peaks in the Bragg spectrum (as indicated by the vertical lines in the figure). 

We have also generated approximate Bragg spectra of a condensate in a TOP trap using the two-state model (see Sec.~\ref{sec:mom}), as well as the simplified version of the two-state model, the independent resonance approach (see Sec.~\ref{sec:features}).  Solving the two-state equation (\ref{two_state_eqn}) is two orders of magnitude faster than solving the full Schr$\rm{\ddot{o}}$dinger equation.  Provided that the validity conditions of Eqs.~(\ref{2_state_validity}) and (\ref{2_state_validity_again}) are satisfied, the results of the two-state model are reasonably accurate, as demonstrated by the dashed line in Fig.~\ref{fig:TOP_spec}.  The main discrepancy between the two-state method and the full $C=0$ calculation is the underestimation of the two strongest resonance peaks.  This occurs because the stationary harmonic trap in the translating frame has been neglected in the two-state approach.  The effect of the harmonic trap is to shift population towards zero momentum thus allowing more condensate population to be scattered into the $n=+1$ momentum order.  The calculation using the independent resonance approach is indicated by the dotted line in Fig.~\ref{fig:TOP_spec} and is almost identical to the Bragg spectra calculated using the two-state approach.  This is because each Bragg resonance within the spectrum is well resolved [refer to Eq.~(\ref{inter_sep})] and, futhermore, the spectrum in Fig.~\ref{fig:TOP_spec} is for exactly eight TOP trap cycles and we note that Eq.~(\ref{inter_off}) holds in that case for TOP trap resonances which are consecutive in frequency space.  

\begin{figure}
\includegraphics[width=8.5cm]{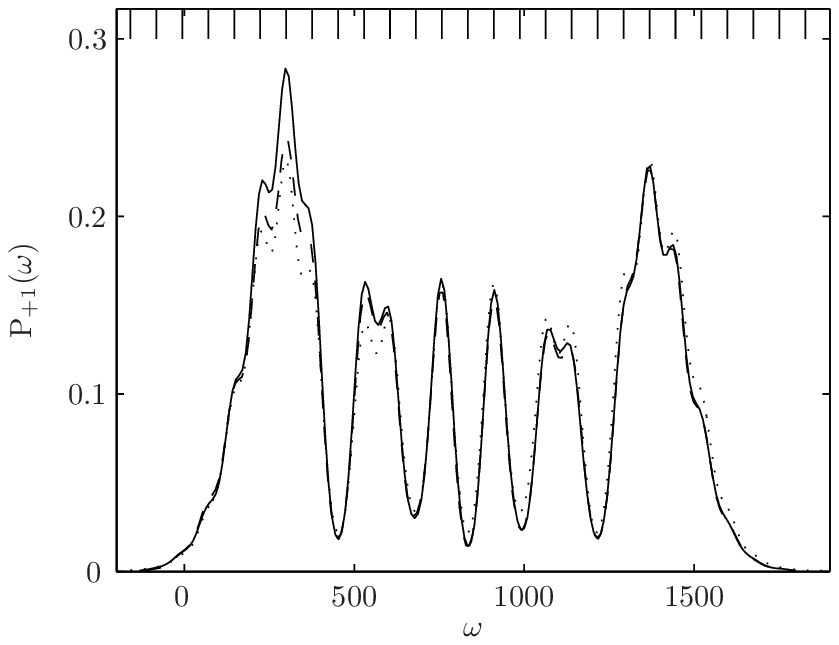}
\caption{\label{fig:TOP_spec_3pt5} Bragg spectrum of a condensate in a TOP trap calculated using the full Schr$\rm{\ddot{o}}$dinger equation (--), the two-state model (- -), and the independent resonance approach ($\cdot \cdot$).  The vertical lines along the top of the figure indicate the positions of the spectral resonances predited by Eq.~(\ref{Veffect}).  Parameters are $U_0=45$, $q=29,$ $r_0=878$, $\Omega =76$, and $t=0.092\pi$ (3.5~TOP trap cycles).}
\end{figure}

Figure~\ref{fig:TOP_spec_3pt5} shows a Bragg spectrum for a condensate in a TOP trap where the bias field rotation frequency is halved compared with the case of Fig.~\ref{fig:TOP_spec}, and the Bragg pulse length is not an integer multiple of that frequency.  
The spectrum is again centred at $\omega=q^2$ and the resonances are separated by the TOP trap rotation frequency, although a number of peaks appear missing (including the central one) because their weight is small in expansion~(\ref{Veffect}).  In contrast to Fig.~\ref{fig:TOP_spec}, the Bragg spectrum of Fig.~\ref{fig:TOP_spec_3pt5} is no longer symmetric about the centre of the first order Bragg resonance, but favours the low frequency resonances due to our initial conditions which yield a negative Doppler term for the first half of the TOP trap rotation.  We observe that the two-state model does exhibit the lack of symmetry present in the full Schr$\rm{\ddot{o}}$dinger equation calculation, while the independent Bragg resonance approach does not.  We also note that the two-state model more accurately approximates the full Schr$\rm{\ddot{o}}$dinger equation calculation than was the case in Fig.~\ref{fig:TOP_spec}.  The reason for this is that neglecting the harmonic trap has less effect at lower scattered fractions.

\subsection{Effect of condensate nonlinearity \label{sec:TOP_nonlinear}}

In this section we present Bragg spectra for a nonlinear condensate in a TOP trap calculated using the Gross-Pitaevskii equation in both one and two dimensions.  In general the Gross-Pitaevskii equation is not separable so we determine the appropriate nonlinear strength in the reduced dimensional system by arranging to have the same Thomas-Fermi radius and chemical potential as in the three-dimensional description.  

\begin{figure}
\includegraphics[width=8.5cm]{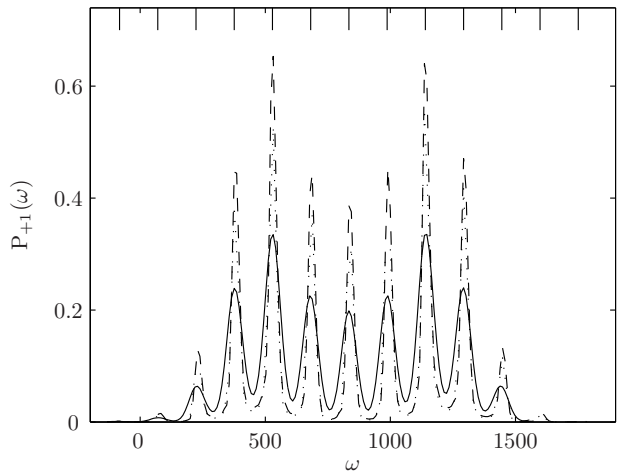}
\caption{\label{fig:nonlinear_8TOP} Bragg spectrum of a condensate in a TOP trap with the parameters of Fig.~\ref{fig:TOP_spec} calculated using the Schr$\rm{\ddot{o}}$dinger equation (--), the one-dimensional Gross-Pitaevskii equation with $C_{1D}=85,$ $\mu=10.08$ (- -), and the two-dimensional Gross-Pitaevskii equation with $C_{2D}=600,$ $\mu=9.85$ ($\cdot \cdot$).  The vertical lines along the top of the figure indicate the positions of the spectral resonances predicted by Eq.~(\ref{Veffect}).}
\end{figure}

Figure~\ref{fig:nonlinear_8TOP} illustrates the Bragg spectra for the same parameters as in Fig.~\ref{fig:TOP_spec} but with $C\neq0$.  The nonlinear strength corresponds to approximately $2 \times 10^4$ atoms in the Otago TOP trap \cite{Wilson03}.  
The results from the one-dimensional Gross-Pitaevskii equation agree well with the Bragg spectrum calculated in two-dimensions, justifying our simplification of the problem to one dimension.  Calculating the two-dimensional Bragg spectra  in Fig.~\ref{fig:TOP_spec} is computationally expensive, requiring approximately $10$~CPU hours on a $3$~GHz Pentium to calculate each point (with 100 points required for the full spectrum).  

We have also reproduced in Fig.~\ref{fig:nonlinear_8TOP} the result from the linear case of Fig.~\ref{fig:TOP_spec}, and comparing the $C=0$ and the $C\neq0$ spectra the effects discussed in Sec.~\ref{sec:dim} are observed.  The nonlinear Bragg spectra exhibit increased peak scattered fractions, reduced resonance width, and a slight increase in the resonant frequencies, with respect to the equivalent $C=0$ system.  The increased resolution in the features of the Bragg spectrum occurs due to the reduced momentum width of the initial state, determined by the form of $\xi (\mathbf{R}).$  It is interesting to note that the two-state model can also exhibit the narrowed resonances if the initial state is taken to be the same numerically calculated ground state as was used for the full nonlinear Gross-Pitaevskii calculation.  

\section{Conclusion}

We have presented methods for identifying the key features of the Bragg spectrum
of a condensate which exhibits solitary-wave behaviour.  In particular such an accelerating condensate becomes stationary in a translating frame and an effective optical potential contains the only explicit time-dependence of the system in that frame.  We have generalized the two-state formalism introduced by Blakie et al.~\cite{Blakie00} and have defined the regime in which the approximate spectra generated using that approach are in good agreement with full calculations.  Bragg spectra of a solitary-wave exhibit multiple resonances and we have identified particular cases where those resonances each act independently of the others, thereby revealing the main features of the Bragg spectrum with minimal calculation.  
  
The Bragg spectrum of a condensate in a TOP trap has been used to illustrate the application of our theory.  We have described the resonance structure in the spectra and in the validity regime of our treatment have demonstrated excellent agreement between our approximate methods and full theoretical calculations.

\acknowledgments{ This work was supported by Marsden Fund 02-PVT-004 and the Foundation for Research Science and Technology (TAD~884). }

\end{document}